\documentstyle[11pt,aaspp4]{article}

\def\iras{{\sl IRAS}}
\def\etal{{\it et~al.}}
\def\vev#1{\left\langle #1\right\rangle}
\def\kms{\ifmmode {{\rm \ km \ s^{-1}}}
\else{$\rm {km \ s^{-1}}$}\fi}
\def\f{{\cal F}}
\def\d{{\rm d}}
\newcommand{\gtsima}{$\; \buildrel > \over \sim \;$}
\newcommand{\ltsima}{$\; \buildrel < \over \sim \;$}
\newcommand{\simgt}{\lower.5ex\hbox{\gtsima}}
\newcommand{\simlt}{\lower.5ex\hbox{\ltsima}}

\lefthead{Koranyi \& Strauss}
\righthead{Testing the Hubble Law}

\begin{document}

\title{Testing the Hubble Law with the {\sl IRAS\/} 1.2 Jy Redshift Survey}

\author{Daniel M. Koranyi}
\authoremail{dkoranyi@cfa.harvard.edu}
\affil{Center for Astrophysics, 60 Garden St., Cambridge, MA 02138}

\and

\author{Michael A. Strauss\altaffilmark{1,2}}
\authoremail{strauss@astro.princeton.edu}
\affil{Institute for Advanced Study, Princeton, NJ 08540}

\altaffiltext{1}{Current address: Dept.~of Astrophysical Sciences, Princeton
University, Princeton, NJ 08544}
\altaffiltext{2}{Alfred P. Sloan Foundation Fellow}

\begin{abstract}
We test and reject the claim of Segal \etal\ (1993) that the
correlation of redshifts and flux densities in a complete sample of
\iras\ galaxies favors a quadratic redshift-distance relation over the
linear Hubble law.  This is done, in effect, by treating the entire
galaxy luminosity function as derived from the 60-$\mu$m 1.2~Jy \iras\
redshift survey of Fisher \etal\ (1995) as a distance indicator;
equivalently, we compare the flux density distribution of galaxies as
a function of redshift with predictions under different
redshift-distance cosmologies, under the assumption of a universal
luminosity function.  This method does not assume a uniform
distribution of galaxies in space.  We find that this test has rather
weak discriminatory power, as argued by Petrosian (1993), and the
differences between models are not as stark as one might expect {\it a
priori}.  Even so, we find that the Hubble law is indeed more strongly
supported by the analysis than is the quadratic redshift-distance
relation.  We identify a bias in the the Segal \etal\ determination of
the luminosity function, which could lead one to mistakenly favor the
quadratic redshift-distance law.  We also present several
complementary analyses of the density field of the sample; the galaxy
density field is found to be close to homogeneous on large scales if
the Hubble law is assumed, while this is not the case with the
quadratic redshift-distance relation.
\end{abstract}

\keywords{}

\section{Introduction}

While the precise value of the Hubble constant $H_0$ is still the
subject of controversy (cf. Jacoby \etal\ 1992), the linearity of the
redshift-distance relation is generally not questioned by most
astronomers.  Since its original announcement by Hubble (1929),
observational evidence has mounted steadily in its 
favor; for recent discussions, see Lauer \& Postman (1992), Peebles
(1993), Riess, Press, \& Kirshner (1996), and Perlmutter \etal\
(1996).  However, Segal and collaborators have persistently argued for
a quadratic redshift-distance relation wherein $z \propto r^p$, where
$z$ is the redshift, $r$ is the distance, and $p=2$ (Segal \etal\
1993, hereafter SNWZ; Segal \& Nicoll 1995, and references therein).
This relation is predicted for the low-redshift regime in the
Chronometric Cosmology developed by Segal (1976).

The greatest difficulty both in establishing the Hubble Law and in
measuring $H_0$ is determining the distances to astronomical objects;
this is often accomplished through the use of standard candles. The
approach of Segal and collaborators in effect is to use the entire
luminosity function of galaxies as a standard candle. If one assumes
that the luminosity function of galaxies is independent of position
and local density, at least after averaging over solid angle in a
large redshift survey, then the comparison of the derived luminosity
function as a function of redshift in principle could be a test of the
assumed redshift-distance relation.  Equivalently, the distribution of
apparent magnitudes as a function of redshift in principle contains
information about the redshift-distance relation (but see Petrosian
1993 and below).  SNWZ choose to examine low-order statistics from the
observed distribution of apparent magnitudes $m$ and redshifts $z$ in
a given sample.  In particular, they calculate the scatter in $m$, and
the slope of the regression of magnitude on log redshift, comparing
the expected values (under the assumption of various values of $p$) to
what is actually observed.  The distribution of apparent magnitudes or
flux densities at a given redshift is independent of the density distribution
of galaxies, and makes no assumption about the large-scale homogeneity
of the universe.  SNWZ have carried out this analysis using the
1.936~Jy redshift survey of galaxies observed with the {\it Infrared
Astronomical Satellite} (\iras) of Strauss \etal\ (1990; 1992). They
found that $p = 1$ (the Hubble law) was strongly rejected by their
analysis, with $p = 2$ (the Lundmark (1925) law) reproducing the
observed magnitude-redshift correlations.

Despite the importance of Segal's claims over the years, there has
been very little response in the literature to this work. Soneira
(1979) calculated the quantity $\langle z | m\rangle$ for the galaxies
in the {\it Reference Catalogue of Bright Galaxies\/} (de Vaucouleurs
\& de Vaucouleurs 1964) and found that $p = 1$ was greatly favored by
the data. However, this analysis requires the assumption of a
homogeneous galaxy distribution, unlike the analysis of SNWZ (cf.,
Nicoll \& Segal 1982). A similar analysis has been carried out by
Shanks, as quoted in Peebles (1993, \S 5); see also Cho\l oniewski
(1995). A more general discussion of the consequences of the quadratic
redshift-distance law is given in Salpeter \& Hoffman (1986). Finally,
recent comparisons of observational data with the predictions of
Chronometric Cosmology at high redshift (which we do {\it not\/}
address in this paper) can be found in Segal \& Nicoll (1986), Wright
(1987; cf., Segal 1987), Cohen \etal\ (1988; cf., Segal 1990), Efron
\& Petrosian (1992), and Segal \& Nicoll (1996).

  Petrosian (1993) argues that given flux densities and redshifts for a
complete sample, it is impossible to separate cosmological effects
(including evolution) from the luminosity function; one must make
additional assumptions about underlying distributions.  He therefore
argues that one cannot use the observed correlations between flux density and
redshift as a test of cosmologies.  The results of this paper are in
accord with this; we find that without additional assumptions about
the underlying density field, the tests carried out by Segal are only
very weak discriminants of cosmological models.

In the present paper, we carry out an analysis similar to that of SNWZ,
using data from a redshift survey of \iras\ galaxies. We discuss the
derivation of the universal luminosity function in \S 2. In
\S 3.1, we follow SNWZ in using the distribution of
redshift and flux density as a function of redshift as a test of
cosmological model. In \S 3.2, we allow ourselves the assumption that the
galaxy distribution is homogeneous on large scales, giving us a
variety of further cosmological tests. We conclude in \S 4.

\section{Derivation of the Luminosity Function}

The luminosity function $\Phi (L)$ of galaxies is the
distribution of galaxies as a function of luminosity:
$\Phi (L) \, dL$ is the mean number of galaxies per unit volume, 
with luminosity between $L$ and $L+dL$.  We will convert from redshift
to distance by writing 
$cz = h_p r^p$, and
arbitrarily adopt values of $h_p=100\,\kms\,{\rm Mpc}^{-p}$. With
these conventions, the luminosity $L$ of  
a galaxy is given by $L = 4\pi (cz/h_p)^{2/p} \nu f$, where $f$ is the
observed flux density at frequency $\nu$. 

Following the notation of Yahil \etal\ (1991) we will find it useful to define
the cumulative luminosity function $\Psi(L)$:
\begin{equation}
\Psi (L) = \int\limits_L^\infty \Phi(L')\, \d L'\qquad ,
\label{eq:Psi} 
\end{equation}
implying that
\begin{equation}
\Phi(L) = -{{\d\Psi} \over {\d L}}\qquad .
\end{equation}
The cumulative luminosity function is closely related to the selection function $\phi(z)$,
which is the fraction of the luminosity function which enters the
sample at a given redshift. For a sample that is flux-density-limited to
$f \geq f_{min}$, we have
\begin{equation}
\phi (z) = { {\Psi [L_{min}(z)]} \over {\Psi [L_{min}(z_s)]} }\qquad ,
\label{eq:selfunct}
\end{equation}
where 
\begin{equation} 
L_{min} \equiv 4\pi (cz/h_p)^{2/p} \nu f_{min}
\label{eq:lmin} 
\end{equation}
is the minimum luminosity detectable at redshift $z$, and
$L_{min}(z_s)$ is a self-imposed lower limit on luminosity,
corresponding to that of a galaxy at the flux density limit placed at
$cz_s = 500 \kms$.

Methods for deriving luminosity functions are reviewed by Binggeli,
Sandage, \& Tammann (1988) and by Strauss \& Willick (1995), and can
be classified as parametric or non-parametric.  The former assume some
implicit functional form for the luminosity function, characterized by
a number of parameters which are adjusted to provide the best fit to
the observational data.  The latter make no assumptions as to the form
of the luminosity function, and are constructed directly from the
data.  However, both approaches assume that the luminosity function is
universal; that is, independent of position or local
density. This assumption of universality is of course a necessary one
if we wish to use the observed luminosity distribution for
cosmological tests (although, as Petrosian 1993 argues, and as we show
below, it is not sufficient).

Here we will derive the luminosity function by maximum-likelihood
estimation, following Nicoll \& Segal (1978; 1980), Sandage, Tammann,
\& Yahil (1979), Nicoll \& Segal (1983), Efstathiou, Ellis, \& Peterson
(1988), and Yahil \etal\ (1991).  If the luminosity function is
universal, then the joint probability density that we have a galaxy of
luminosity $L_i$ with
redshift $z_i$ is a separable function of these two variables:
\begin{equation} 
\f(L_i,z_i) \,\d L\, \d z = \Phi(L_i) \rho(z_i) \,\d L\, \d z \qquad,
\label{eq:joint-prob} 
\end{equation}
where $\rho(z)$ is the local density of galaxies at redshift $z$.  
The conditional probability density $\f(L_i|z_i)$ that the galaxy have luminosity $L_i$,
given its redshift $z_i$, is then given by the joint probability,
divided by the integral of
the joint probability over all possible luminosities in the sample
at that redshift, given the flux density limit.  That is,
\begin{equation}
\f(L_i|z_i) = {{\f(L_i, z_i)} \over \int_{L_{min}(z_i)}^\infty \f(L,
z_i)\, \d L} = {\Phi(L_i) \over
      {\!\!\int\limits_{L_{min}(z_i)}^{\infty}\!\! \Phi(L)\, \d L} }\qquad .
\label{eq:indivlike}
\end{equation}
Note that the density field $\rho(z)$ has dropped out of this
equation. 
The quantity $L_{min}$ is defined in Eq.~(\ref{eq:lmin}); the
$p$-dependence enters the analysis when calculating the luminosity of
each galaxy from its redshift and flux density. 

We then maximize the likelihood of observing the entire sample: 
\begin{equation}
{\cal L} =\prod_i {\f(L_i|z_i)} \label{eq:totallike}\qquad. 
\end{equation}
For computational convenience, we instead {\it minimize} the
{\it negative logarithm} of the likelihood, where
\begin{equation}
-\ln {\cal L}  = - \sum_i \ln \f(L_i|z_i)\qquad.
\label{eq:loglike} 
\end{equation}
An attempt was made to include $p$ as one of the parameters with
respect to which $\cal L$ was maximized, but this proved impossible;
$\cal L$ turns out to be a monotonically increasing function of $p$.

The maximum-likelihood approach does not depend on the assumption that
galaxies are distributed uniformly, since $\rho(z)$ dropped out of
Eq.~(\ref{eq:indivlike}).  However, this 
requires that the mean spatial density of galaxies in the sample 
be calculated by other means.  There is a variety of such density estimators
available (Davis \& Huchra 1982); the one
chosen here is
\begin{equation}
n_1 \equiv \Psi\left[L_{min}(z_s)\right] = V^{-1} \sum_i {1 \over \phi(z_i)}\qquad,
\label{eq:n1} 
\end{equation}
where the sum is over all galaxies in the sample volume $V$, and
$\phi(z_i)$ is the value of the selection function at the redshift of the
$i^{\rm th}$ galaxy, as in Eq.~(\ref{eq:selfunct}).

The data set consists of 5313 pairs of 60 $\mu$m \iras\ flux densities
and redshifts above a flux density limit of 1.2~Jy, taken from Strauss
\etal\ (1992) and Fisher \etal\ (1995).  The recession velocities are
corrected for the motion of the Sun with respect to the barycenter of
the Local Group following Yahil, Tammann, \& Sandage (1977).  Of these
galaxies, 4218 have redshifts in the range $500 \leq cz \leq 12000$ km
s$^{-1}$; only these were used in fitting the luminosity function
since the sample may be incomplete at much higher redshifts (Fisher \etal\
1992), and at low redshifts the local motions are strongly affected by
peculiar velocities that dominate the local Hubble expansion.

\subsection{Parameterized derivation}

Various parameterizations of the luminosity function are discussed by Strauss (1989) and
Saunders \etal\ (1990).
We follow Yahil \etal\ (1991), in parameterizing the cumulative luminosity function by
\begin{equation}
\Psi(L) = C \left( {L\over L_*} \right) ^{-\alpha}
            \left( 1 + {L \over L_*} \right) ^{-\beta}\qquad,
\label{eq:Yahil} 
\end{equation}
so that the differential luminosity function is given by
\begin{equation}
  \Phi(L) = -{\d\Psi(L) \over \d L} =
  \left( {\alpha \over L} + {\beta \over {L_*+L}} \right) \Psi(L)\qquad,
\label{eq:Yahil-diff} 
\end{equation}
and the minimization is performed with respect to the two
dimensionless parameters $\alpha$ and $\beta$ and the characteristic
luminosity $L_*$, using the routine {\tt mrqmin} from Press
\etal\ (1992).  The optimal values of the parameters for $p=1,2,3$
are tabulated in Table~\ref{table:yahil123parms}; the
case $p=3$ will be included in some of the tests we present for 
methodological perspective.

\placetable{table:yahil123parms}

For graphical purposes, it is often convenient to work with the
distribution per volume per {\it log} luminosity, given by
\begin{equation}
   \hat{\Phi} (L) \equiv -{ {\d\Psi} \over {\d({\rm log_{10}} L)} } =
   {1 \over { {\rm log_{10}} e} } L \Phi (L)\qquad.
\end{equation}
We shall use this form in making plots of the luminosity function.

\subsection{Non-parameterized derivation}

One drawback of assuming a parameterized form of the luminosity
function is that it constrains the luminosity function to have a
certain functional form, which may not be general enough to accurately
reflect the actual luminosity function.
A non-parametric approach to determining the luminosity function does
not suffer from this limitation.

We model the differential luminosity function as being
piecewise-constant over $n$ bins evenly spaced in $\log_{10} L$ from
the minimum to maximum luminosities seen in the sample, and then treat
the value of the function on these bins as the $n$ parameters with
respect to which the likelihood (defined exactly as before) is to be
maximized.  This avoids any implicit assumptions of the functional
form that the luminosity function should have, but also sacrifices any
requirements as to continuity and smoothness that physical intuition
suggests should be satisfied.  This approach was first suggested by
Nicoll \& Segal (1980; 1983), and was reinvented by Efstathiou \etal\
(1988). Rather than attempting to search for a minimum in some high
$n$-dimensional parameter space, the technique is to converge
iteratively to the optimal step values.  The interested reader is
referred to the Appendix or Efstathiou \etal\ (1988) for details.

There are several drawbacks to the non-parametric luminosity function
method.  The most obvious is that the resultant luminosity function is
discontinuous; this can be overcome by interpolating linearly between
the centers of the bins; the details are set forth in the Appendix. We
in fact use this interpolating method in what follows below. More
serious is the effect illustrated below; a strong sensitivity of the
derived luminosity function to bin size.  

  The ROBUST method of SNWZ calculates the luminosity function in two
steps.  For a given sample with maximum redshift $z_{max}$, there is a
luminosity $L_{min}(z_{max}) = 4\,\pi \left(cz_{max}/h_p\right)^{2/p}\nu f_{min}$
above which a galaxy can be found anywhere in the volume of the
survey, so $\Phi(L)$ in this luminosity range is simply proportional
to the number of galaxies at each value of $L$.  For lower
luminosities, however, the volume in which a galaxy could be found is
an increasing function of the luminosity, requiring further
calculation to determine the luminosity function.  Nicoll \& Segal
(1983) and SNWZ argue that the number of bins $n_{ll}$ used {\it in
the lower-luminosity range alone\/} should be set the same when
comparing different cosmologies, in order to avoid giving any one
cosmology extra degrees of freedom.  Clearly, $n_{ll} < n$ by
definition.

\begin{figure}
\plotone{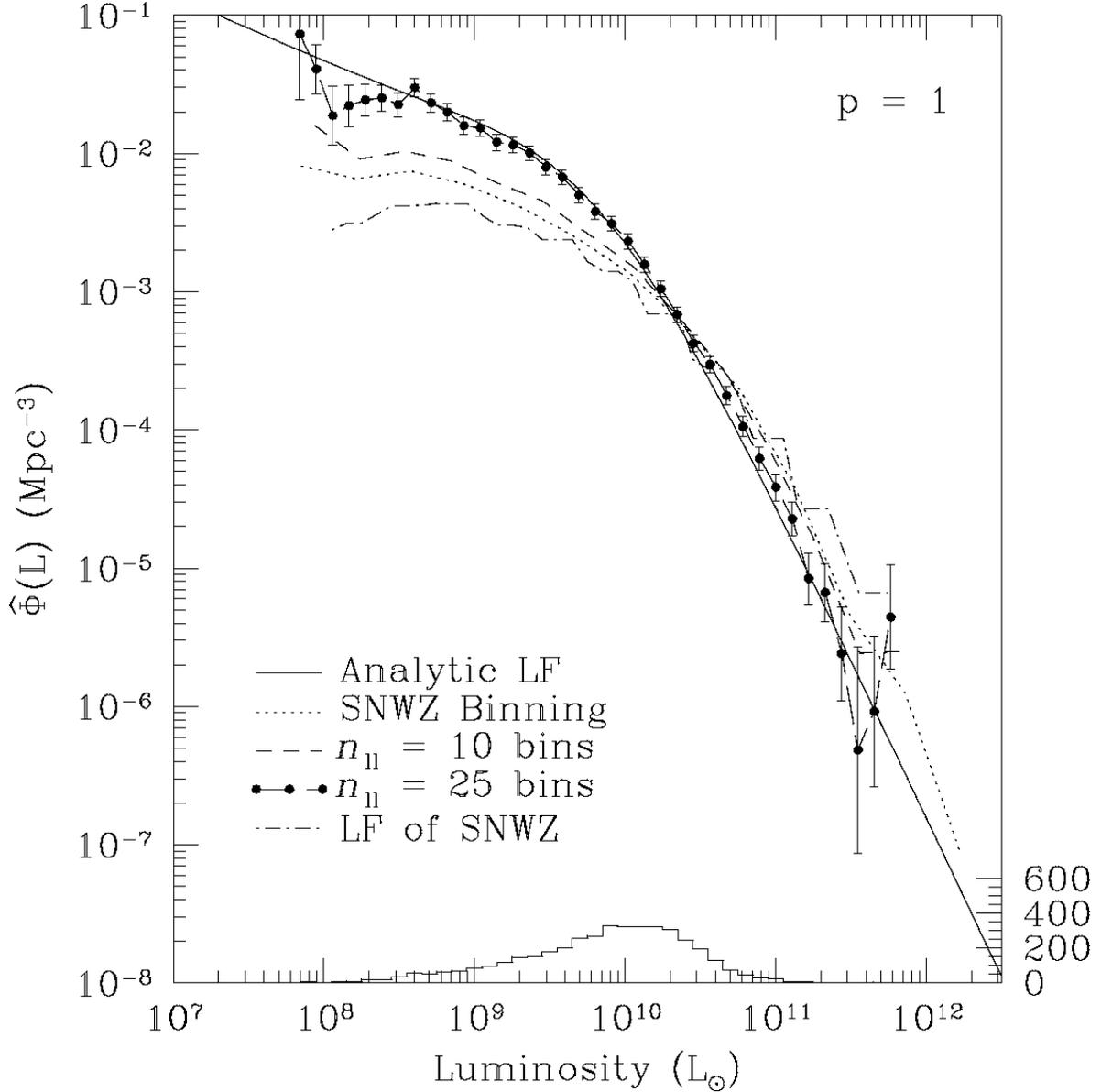}
\caption[]{Plots of the luminosity function derived by both
parametric and non-parametric methods for $p=1$; the solid curve is
parametric, and three non-parametric luminosity functions are shown,
with different binning in $\log L$.
Error bars are suppressed except for the luminosity function with the
finest binning, which matches the parameterized luminosity function
quite closely.  The luminosity function of SNWZ is plotted for
comparison.  The histogram at the bottom indicates the luminosity
distribution of the galaxies in the
sample. \protect\label{fig:comparephi_p1}}
\end{figure}

\begin{figure}
\plotone{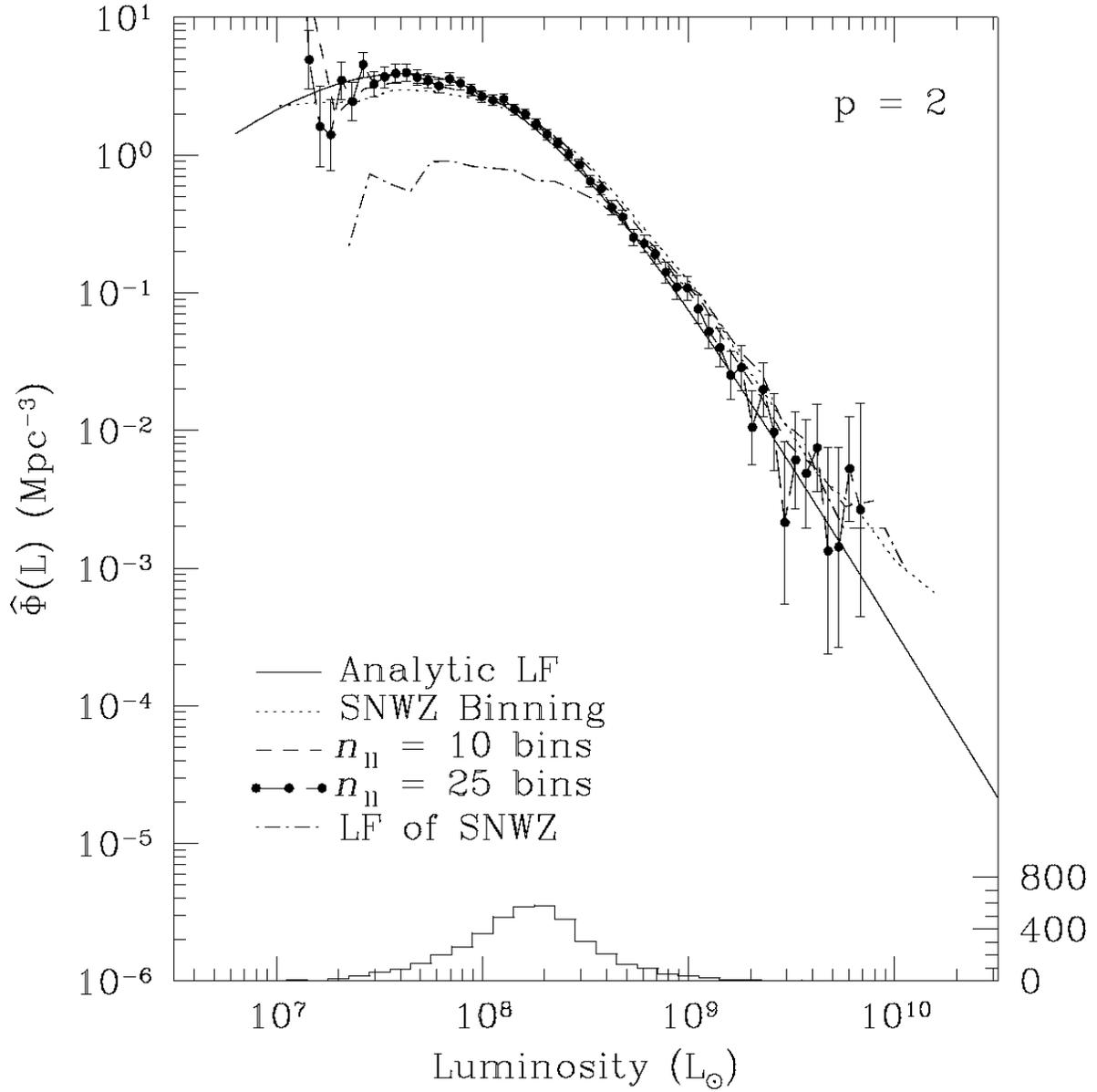}
\caption[]{
As in Fig.~\ref{fig:comparephi_p1}, but
for $p=2$. \protect\label{fig:comparephi_p2}}
\end{figure}

With this in mind, Figs.~\ref{fig:comparephi_p1} and \ref{fig:comparephi_p2}
show the resultant luminosity functions derived under the assumption
of $p=1$ and $p=2$, respectively.  We plot the parameterized
luminosity function (Eq.~\ref{eq:Yahil-diff}, using the parameters of
Table~\ref{table:yahil123parms}) as a solid curve, and superimpose the
luminosity functions derived using the nonparametric method for
several different binnings: the bin size used by SNWZ (in $\log L$),
$n_{ll} = 10$ as recommended by SNWZ, and $n_{ll} =
25$\footnote{Because the range of luminosities below
$L_{min}(z_{max})$ is dependent on the sample and the assumed value of
$z_{max}$, our $n_{ll} = 10$ case has a different bin size than that
of SNWZ.}.  We plot error bars only for the non-parametric luminosity
function that most closely approaches the parametric one.  The bin
sizes and number of bins for these and other luminosity functions are
tabulated in Table~\ref{table:segalstats} below.  We also plot the
luminosity functions of SNWZ for comparison (normalized using
Eq.~\ref{eq:n1}). Different bin sizes result in appreciable
differences in the faint end of the luminosity function; in
particular, the fewer the number of bins, the more the faint end of
the luminosity function is attenuated. 

At the
bottom of these plots we have placed histograms indicating the
distribution of the luminosities from which our luminosity functions
were computed; note that the luminosity distribution of the galaxies
in the sample narrows significantly with increasing $p$.

For $p = 1$, note that the luminosity function we find using the SNWZ
binning is close to the luminosity function found by SNWZ themselves.
However, this bin size is large enough to strongly bias the luminosity
function at the faint end; the luminosity function does not become
stable until the number of bins approaches $n = 20$ or more
(corresponding to $n_{ll} = 15$).  Thereafter, the luminosity function
found by the non-parametric method, which is in principle free to
assume any shape at all, is in good agreement with the
parametric luminosity function.  For $p = 2$, the sensitivity to
binning seems to be much less severe, and the luminosity functions we
find for all binnings are in very good agreement.  In this case, we
were able to reproduce the SNWZ luminosity function only with a very
small number of bins $n_{ll} = 5$, corresponding to $n = 10$ (not
shown). 


The data set used by SNWZ consisted of the brighter half of the
present one, flux density-limited at 1.936~Jy at 60~$\mu$m. This makes
little difference; we find that the luminosity function from different
subsets in flux density is quite robust (Koranyi 1993), as long as
$n_{ll}$ is large enough. 

 The biasing in the luminosity function due to binning is more severe
for $p = 1$ than for $p = 2$, and therefore statistics for different
values of $p$ using these luminosity functions can give misleading
results if the binning used is overly coarse.  Indeed, we now show that
with the bin size recommended by SNWZ, $p = 2$
is indeed a better fit to the data than is $p = 1$, while with finer
binning, $p = 1$ is preferred. 


\placefigure{fig:comparephi_p1}
\placefigure{fig:comparephi_p2}

\section{Comparison of Different Cosmologies}

Having derived a luminosity function for the sample under the
assumption that the density and luminosity distributions of the sample
are separable, we are now in a position to test the relative merits of
the Hubble and Lundmark laws.  We can test for robustness of the
luminosity function, and also for self-consistency in the predictive
powers of the luminosity function under the assumption of various
power law cosmologies.  These comparisons separate naturally into
those that do and do not depend on the assumption that the galaxy
distribution is uniform on large scales. We start with the latter, following SNWZ.

\subsection{Density-independent comparisons}

\begin{figure}
\plotone{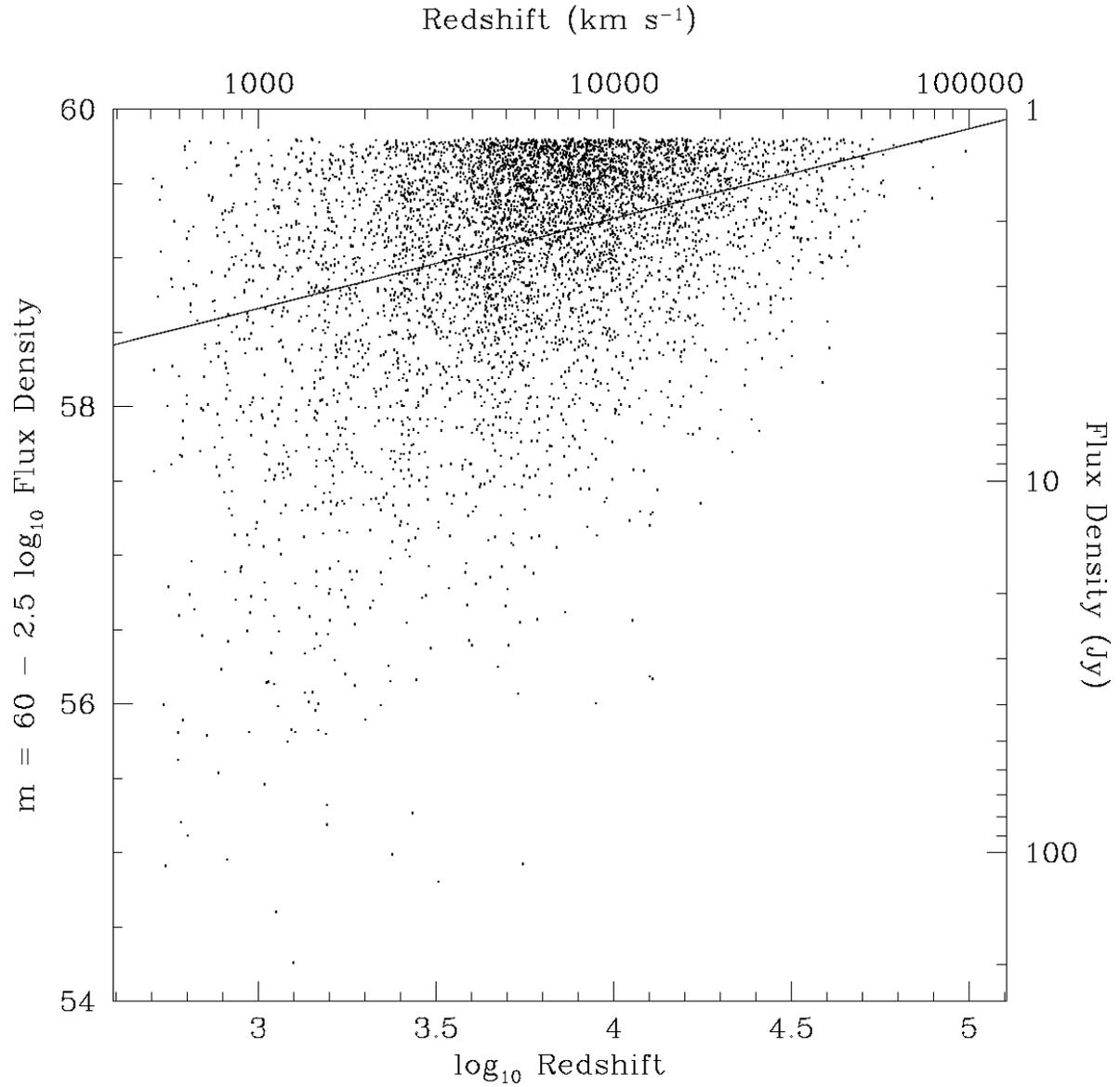}
\caption[]{The observed distribution of apparent magnitudes $m$ as a
function of $\log_{10} cz$.  The line is the best-fit regression for
those galaxies with $2000 < cz < 20,000 \kms$.}
\protect\label{fig:mag-redshift}
\end{figure}

SNWZ argue strongly that it is important to distinguish between
cosmological tests that involve quantities which depend on the
cosmology itself (such as luminosity, or the luminosity function), and
those that are pure observables (such as flux density, or equivalently
magnitude, and redshift). Moreover, they develop statistical tests
that do not depend on the assumption that the distribution of galaxies
is uniform in space. The statistics on which they put the greatest
weight are $s$, the standard deviation of apparent magnitudes in a sample, and
$\beta$, the slope of the regression of the apparent magnitudes on log
redshift (the notation is that of SNWZ).  Before calculating these
statistics, it is useful to examine Fig.~\ref{fig:mag-redshift}, which is
the observed relation between apparent magnitudes (defined here,
following SNWZ, as $m = 60 - 2.5 \log_{10} f$) and $\log_{10}
cz$.  The correlation between these two quantities is not very
strong. The line shown is the regression line of $m$ on $\log_{10}
cz$ for the subsample of galaxies with $2000 < cz < 20,000 \kms$; we
have found that these statistics are very sensitive to the exact
lower-redshift cutoff at redshifts below 2000 \kms. 
Table~\ref{table:segalstats} gives the observed values of $s$ and
$\beta$.  SNWZ calculate the expected values of these statistics using
Monte-Carlo simulations.  Here we will do so analytically. 

\placefigure{fig:mag-redshift}
\placetable{table:segalstats}


Given a model for the luminosity function and a value of $p$, one can
calculate the probability distribution function of flux density $f$ of
galaxy $i$, given its redshift $z_i$. The distribution function for a
single galaxy is given by the luminosity function, normalized
appropriately, following Eq.~(\ref{eq:indivlike}):
\begin{equation}
\f(f|z_i)={{{\d L \over \d f}\, \Phi(L)}\over
               {\int_{L_{min}(z_i)}^\infty \!\!\Phi(L')\d L'}}
                \propto {{\Phi(L)}\over{\phi(z_i)}}\quad,
\label{eq:lumdist}	
\end{equation}
where $\phi(z)$ is the selection function, 
$L = 4\,\pi \nu (cz_i/h_p)^{2/p} f$, and $L_{min}$ was defined above
in Eq.~(\ref{eq:lmin}).  From this distribution function,
we can easily  calculate the moments of the apparent magnitudes, and
therefore the scatter $s$.  Similarly, we can calculate the
expectation value of $m$ for each value of $z$, from which $\beta$
follows directly.  Note that both these statistics depend on the
observed distribution of redshifts, and cannot be considered
independent of this.  

  Table~\ref{table:segalstats} shows the predicted results for $p = 1,
2$, and 3 for various binnings for the subsample of galaxies with
$2000 < cz < 20,000 \kms$.  It is clear that the results are
quite sensitive to the details of the binning.  Let us start by
concentrating on the results using the binning of SNWZ (the first row
for each value of $p$).  Both $p = 2$ and $p = 3$ do a much better job
of predicting the observed apparent magnitude scatter $s$ than does $p
= 1$.  We interpret this to be due to the fact that the $p = 1$
luminosity function is much more biased at this coarse binning than
that of $p = 2$.  We find similar results at $n_{ll} = 10$, as recommended
by SNWZ.  None of the models do a good job of reproducing the slope of
regression, $\beta$, with this binning, with $p = 1$ overpredicting
the observed value of $\beta$ by as much as $p = 2$ underpredicts
it. It is clear from Fig.~\ref{fig:mag-redshift}, however, that a
linear fit to the magnitude-log redshift scatter diagram is a rather
poor way to model the data, and that the results may be quite
sensitive to a small number of outlying points. 

  With finer binning ($n_{ll} = 25$), and with the parametrized
luminosity function, the predictions for $s$ are
essentially {\it independent\/} of $p$.  As Petrosian (1993) first
argued, and as we conclude below, without additional assumptions, it
is very difficult to distinguish cosmologies from redshift and
magnitude data alone.  However, note that the $p = 1$ does predict the
correct value of $\beta$ with this larger number of bins (the
parametrized model does less well), while $p = 2$ and $p = 3$ fail
quite badly.  We describe the $\chi^2$ column in this table below. 

We can ask more of the data than simply these low-order
statistics. Indeed, Eq.~(\ref{eq:lumdist}) gives a prediction for the
flux density distribution of galaxies as a function of redshift; we can
compare this directly with what is observed.  That is, the sum of
these distribution functions over some subsample of a redshift survey
can be compared with the {\it observed\/} flux density distribution
as an {\it a posteriori\/} test of the luminosity function (Sandage
\etal\ 1979; Yahil \etal\ 1991; Strauss \& Willick 1995), or, in the
present application, of the cosmology assumed.  In other words, the
predicted distribution of galaxies of a given flux density $f$ is
\begin{equation}
N(f)=\sum_i{\f(f|z_i)}\qquad,
\end{equation}
where the summation is over all galaxies in a particular (sub)sample.
The predicted and actual values in each bin can be compared with the
$\chi^2$ statistic, yielding a measure of how similar
the two distributions in fact are.

%

\begin{figure}
\plotone{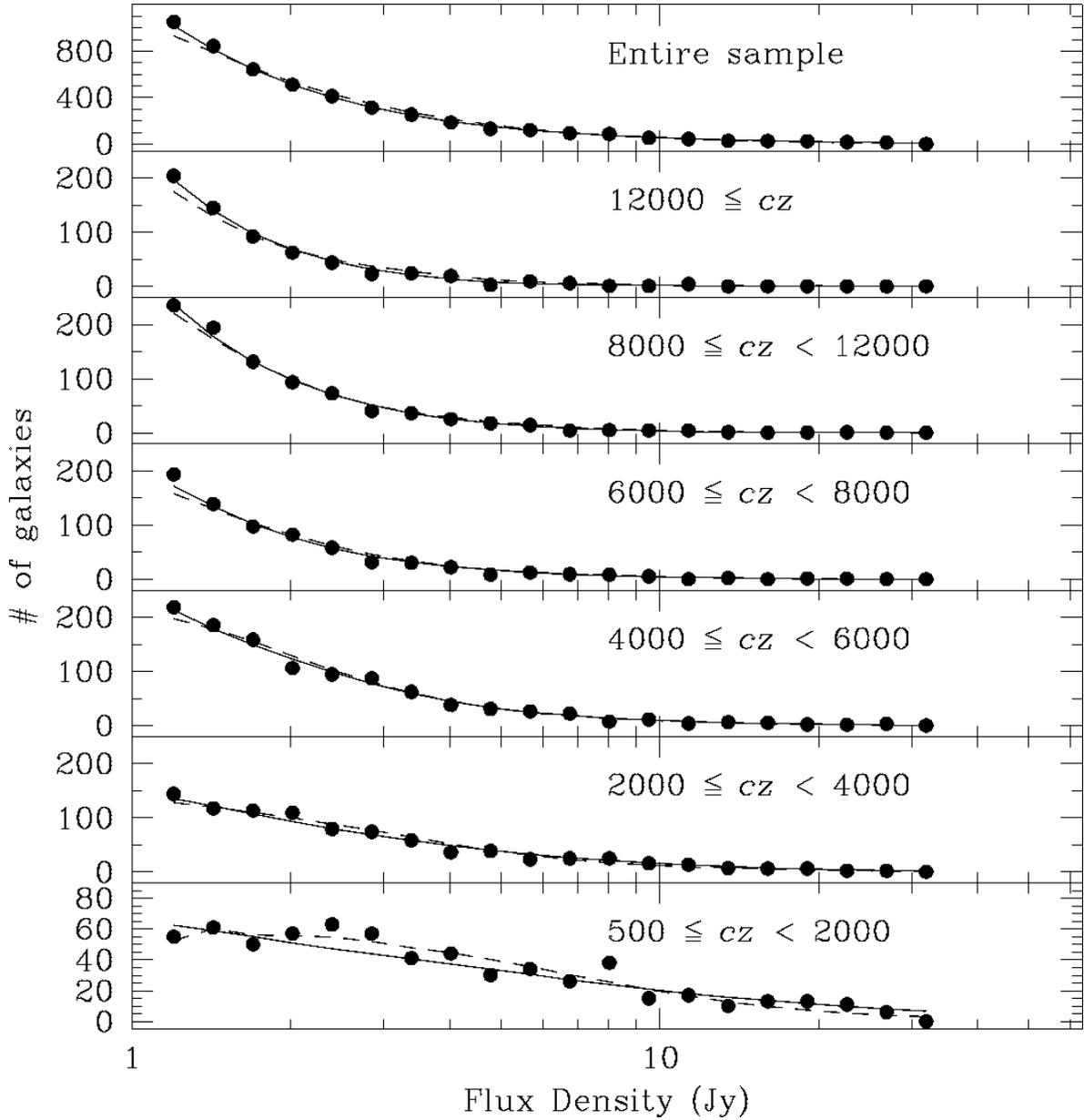}
\caption[]{Observed (points) and expected (curves) distribution
of galaxies in flux density at different redshift ranges (labeled in
\kms).   The solid curves are for $p = 1$, and the dashed curves for
$p = 2$. 
Topmost box is for the entire sample.  Note the virtual
indistinguishability of the predictions for the two power laws; this
shows that the flux density (or equivalently apparent magnitude)
distribution has little discriminatory power.  Predictions are based
on the nonparametric luminosity function fitted over the range
500--12000 km s$^{-1}$, with $n_{ll} = 25$. \protect\label{fig:fluxdist}}
\end{figure}

Fig.~\ref{fig:fluxdist} shows the results for the full sample of galaxies
($500 < cz < 20,000 \kms$), as well as for a variety of redshift
subsamples.  The observed distribution in flux density is indicated by
the dots, and the predictions for $p=1$ and $p=2$, using the
luminosity functions with $n_{ll} = 25$ are indicated by the solid and
dashed curves, respectively.  As expected for a flux density-limited
sample, the distributions peak strongly towards the flux density limit
of 1.2~Jy, with a tail of higher flux density observations; since this
must be the case regardless of cosmology, we are not surprised that
there are no gross morphological differences between the flux density
distributions for the competing cosmologies.  Indeed, the difference
between the predicted curves for $p=1$ and $p=2$ is very small.
Following Yahil \etal\ (1991), we calculate the $\chi^2$ statistic of
the difference between the predicted and observed curves, using
Poisson error bars and summing only over bins with five or more
galaxies.  The results are tabulated in Table~\ref{table:fluxchi2}.  The fit
with $p=1$ is acceptable in all bins.  $p=2$ fares rather worse;
although it is acceptable in several of the redshift bins, the fit for
the full sample is unacceptable. For $p=3$ (not shown in the figure)
the fit in most redshift bins is unacceptable.

  The quantity $\nu$ in Table~\ref{table:fluxchi2} is the
number of bins of flux density in which the
comparison is done.  One might argue that this number should be
reduced by the number of free parameters in the luminosity function
(25, in these cases!).  We make two points here: first, we have found
qualitatively very similar results to those presented here when we use
the parameterized luminosity function of Eq.~(\ref{eq:Yahil-diff}),
which uses only three parameters (Table~\ref{table:yahil123parms}).  Second,
note that the fit to the luminosity function is done for galaxies with
redshifts between 500 and 12,000 \kms; for $p = 1$, this fit remains
good for galaxies between 12,000 and 20,000 \kms, while for $p = 2$
and $p = 3$, the fit is unacceptable in this range.

\begin{figure}
\plotone{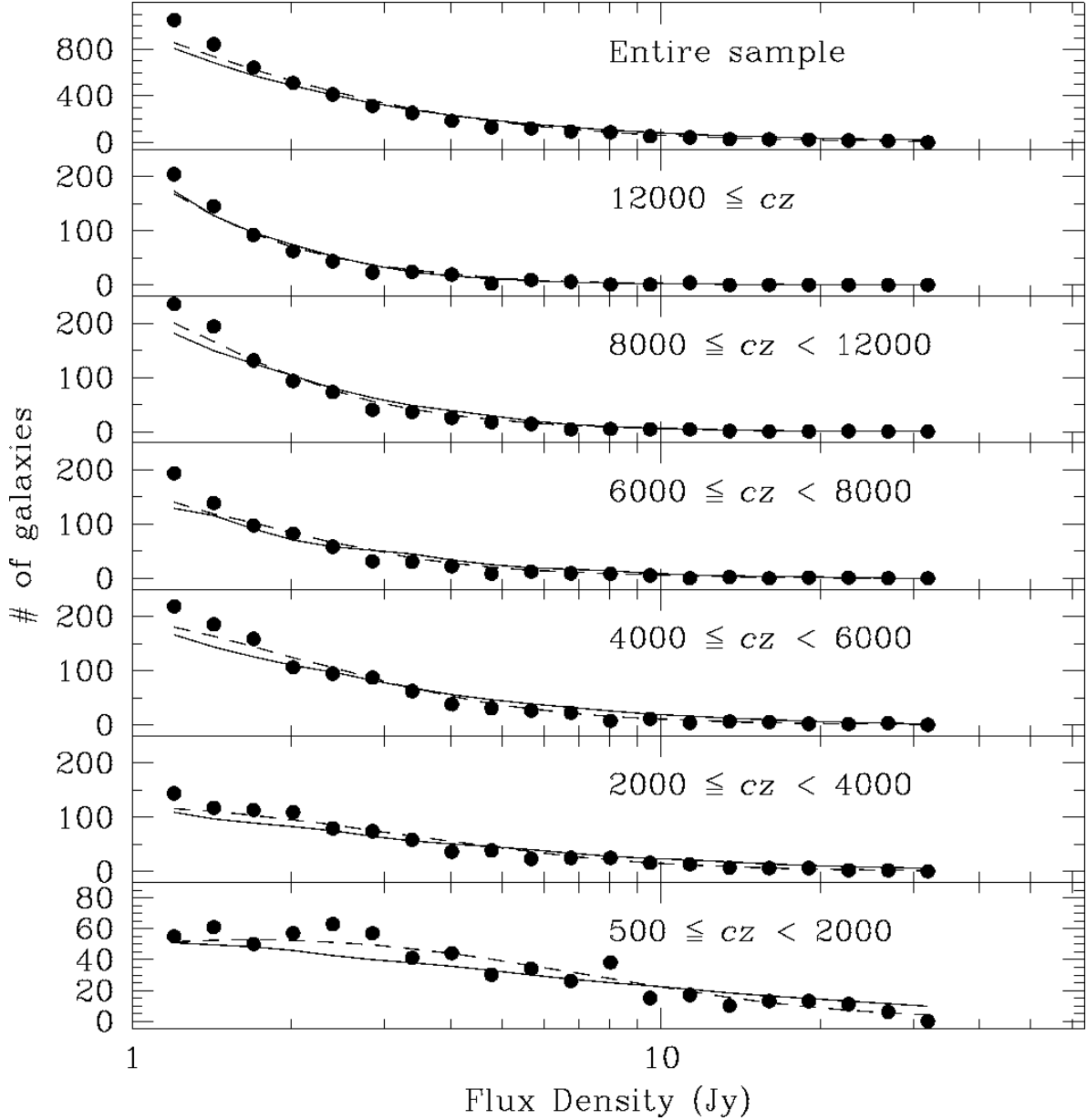}
\caption[]{As in Fig.~\ref{fig:fluxdist}, but the luminosity functions
are calculated with the SNWZ binning for both $p = 1$ and $p = 2$.  Note
that neither curve is a particularly good fit to the data, especially
at the faint end, although $p
= 2$ does somewhat better. \protect\label{fig:fluxdist_segal}}
\end{figure}

\placefigure{fig:fluxdist}
\placefigure{fig:fluxdist_segal}
\placetable{table:fluxchi2}
\placetable{table:fluxchi2_segalstep}

Fig.~\ref{fig:fluxdist_segal} repeats this exercise using the SNWZ
binning.  The fits are not nearly as good as before, especially 
at the lowest luminosities.  This is quantified in the $\chi^2$
statistics tabulated in Table~\ref{table:fluxchi2_segalstep}.  However,
although no value of $p$ is acceptable, $p = 2$ is much preferred over
$p = 1$, especially for the full sample.  We saw in
Figs.~\ref{fig:comparephi_p1} and \ref{fig:comparephi_p2} that $n_{ll} = 10$
corresponds to a bin size which gives a strongly biased luminosity
function for $p = 1$, but that the bias was much less severe for $p =
2$.  

  We explore the effect of binning further in Table~\ref{table:segalstats}.
The $\chi^2$ column gives the $\chi^2$ values (for $\nu = 19$) for the
flux density comparison for a variety of binnings, for the sample with
$2000 < cz < 20,000 \kms$.  By this statistic, $p = 1$ is strongly
ruled out with $n_{ll} = 10$ or with the SNWZ binning (doing much more
poorly than even $p = 3$), while with either $n_{ll} = 25$ or with the
parametrized luminosity function, $p = 1$ gives acceptable results. 

  We believe that this
is the origin of the claims of SNWZ.  With proper determination of the
luminosity function, the distribution function of flux densities given the
redshifts can be predicted almost as well with $p = 2$ as with $p =
1$; it is very insensitive to cosmology, as was pointed out originally
by Petrosian (1993).  However, the luminosity function is biased if
one chooses too low a value of $n_{ll}$, and this effect is more
severe for $p = 1$ than for $p = 2$, causing one to erroneously
conclude that $p = 2$ is preferred by the data.

  Our conclusion from this discussion, mirroring that
of Petrosian (1993), is that {\it any\/} statistic derived from the
distribution of flux densities and redshifts will be able to match the data
roughly equally well for $p = 1$ or $p = 2$, without making further
assumptions about the distributions.  

  One such assumption we could make is that the universe approaches
homogeneity on the largest scales (the Cosmological Principle; cf.,
Peebles 1993).  With such an assumption, we can derive further
statistics that do allow a sharp distinction between different values
of $p$.

\subsection{Density-dependent comparisons}

\begin{figure}
\plotone{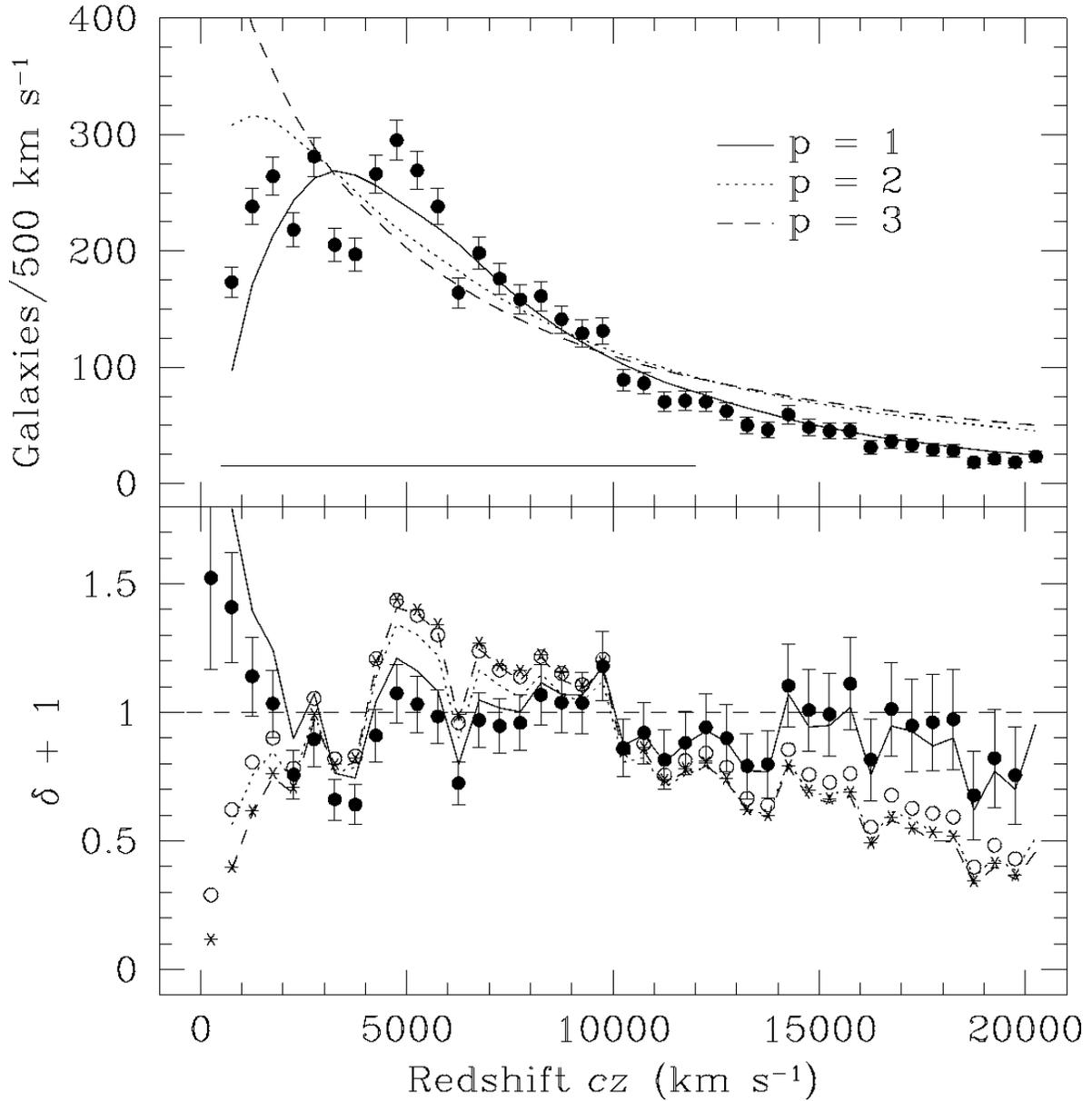}
\caption[]{The upper panel shows the observed (points) and predicted
(curves) distribution of galaxies with redshift for $p=1,2,3$.  The
bottom panel shows the ratio of observed to predicted counts per bin;
the predictions are based on the nonparametric luminosity function.
The bar in the top panel marks the range of redshifts used in fitting
the luminosity function.  The points in the lower panel show the
density field calculated nonparametrically and independently of the
luminosity function, following Saunders \etal\ (1990).  Solid points
are for $p = 1$, open circles for $p = 2$, and stars are for $p = 3$.
Error bars are shown only for $p = 1$ to keep the figure from getting
too cluttered; however, the errors are quite insensitive to the value
of $p$.  \protect\label{fig:zdist}}
\end{figure}

Under the homogeneous approximation, one can
predict the distribution of galaxies with redshift, given the
luminosity function and a value of $p$.  In a shell of thickness
$\Delta z$ at a redshift $z$, one expects there to be 
\begin{equation}
N(z) = {\omega \over p} n_1 \phi(z) \left({{cz}\over{h_p}}\right)^{3/p 
- 1} \, \Delta z
\label{eq:n_of_z}  
\end{equation}
galaxies, where $\omega$ is the solid angle covered by the
sample (11.06 ster in our case; cf. Strauss \etal\ 1990). These
predictions for various $p$ are shown in the upper panel 
of Fig.~\ref{fig:zdist}, along with the observed distribution with
redshift (we use $n_{ll} = 25$ throughout this section).  The lower
panel shows the ratio of the observed to predicted counts (which is
the fractional density of the shell relative to the mean over the
sample).  The $p=1$ curve agrees closely with the observed
distribution, although the agreement is not perfect.  The effects of
the Local Supercluster at $cz \simeq 1500$ km s$^{-1}$ and of the
Great Attractor and Pisces-Perseus regions at $cz \simeq 4500$ km
s$^{-1}$ are visible as overdensities, even when averaged over almost
the entire sky.  At very large redshifts, the $p=1$ curve slightly
overestimates the observed distribution, probably because of
incompleteness of the sample at high redshifts (Fisher \etal\ 1992).
However, if either the $p=2$ or $p=3$ model were correct, one would
have to argue that as one looks to higher redshifts in the universe,
the inhomogeneities grow.  The galaxy density plotted in the lower
panel would need to be a strong function of redshift: low nearby,
rising rapidly to a maximum at 5000 -- 10,000 km s$^{-1}$, and then
dropping by a factor of two thereafter.  This would violate the
Cosmological Principle, and indeed, if such massive structures were
common in the universe, we would see them reflected in the angular
correlation function of faint galaxies ({\it e.g.} Maddox \etal\
1990).

\placefigure{fig:zdist}

Indeed, one can calculate the distribution of densities on shells as a
function of redshift, without any calculation of the luminosity
function at all (Saunders \etal\ 1990).  In \S~2 above, we calculated
the distribution function of luminosities conditioned on the redshifts
$\f(L_i|z_i)$.  Here we calculate the distribution of redshifts
conditioned on the luminosities $\f(z_i | L_i)$, which is given by:
\begin{equation} 
\f(z_i | L_i) = {\f(L_i, z_i) \over {\int \d z\, {\d V \over \d z}\,
\f(L_i, z)}} = {\rho(z_i) \over 4\,\pi \int_0^{z_{max, i}} \d z\, z^{3/p - 1} \rho(z)}
\qquad,
\label{eq:z-given-L} 
\end{equation}
where
\begin{equation} 
z_{max, i} = {h_p \over c} \left({L_i \over 4\,\pi \nu f_{min}}
\right)^{p/2}\qquad. 
\label{eq:zmax} 
\end{equation}
When we conditioned on redshift (Eq.~\ref{eq:indivlike}), the density
distribution dropped out of the expression, while here, the luminosity
function drops out.  We can now maximize the likelihood with respect
to the density field defined at a series of steps, exactly as we did
with the luminosity function\footnote{We will not, however, apply the
additional complication of interpolating the density field between
bins as we do in the Appendix for the luminosity function.}, using the
iterative technique described by Efstathiou \etal\ (1988).  The
results are shown as points in the lower panel of Fig.~\ref{fig:zdist}.
The solid points are for $p = 1$, the open circles for $p = 2$, and
the stars for $p = 3$.  Error bars are given only for $p = 1$ to keep
the figure from getting overly crowded; the error bars are in fact
quite insensitive to the value of $p$.  The agreement between this
density field and that given by the curves, which depends on the
calculation of the luminosity function, is striking; this shows us
that both the luminosity function and density field calculations are
robust. 

\begin{figure}
\plotone{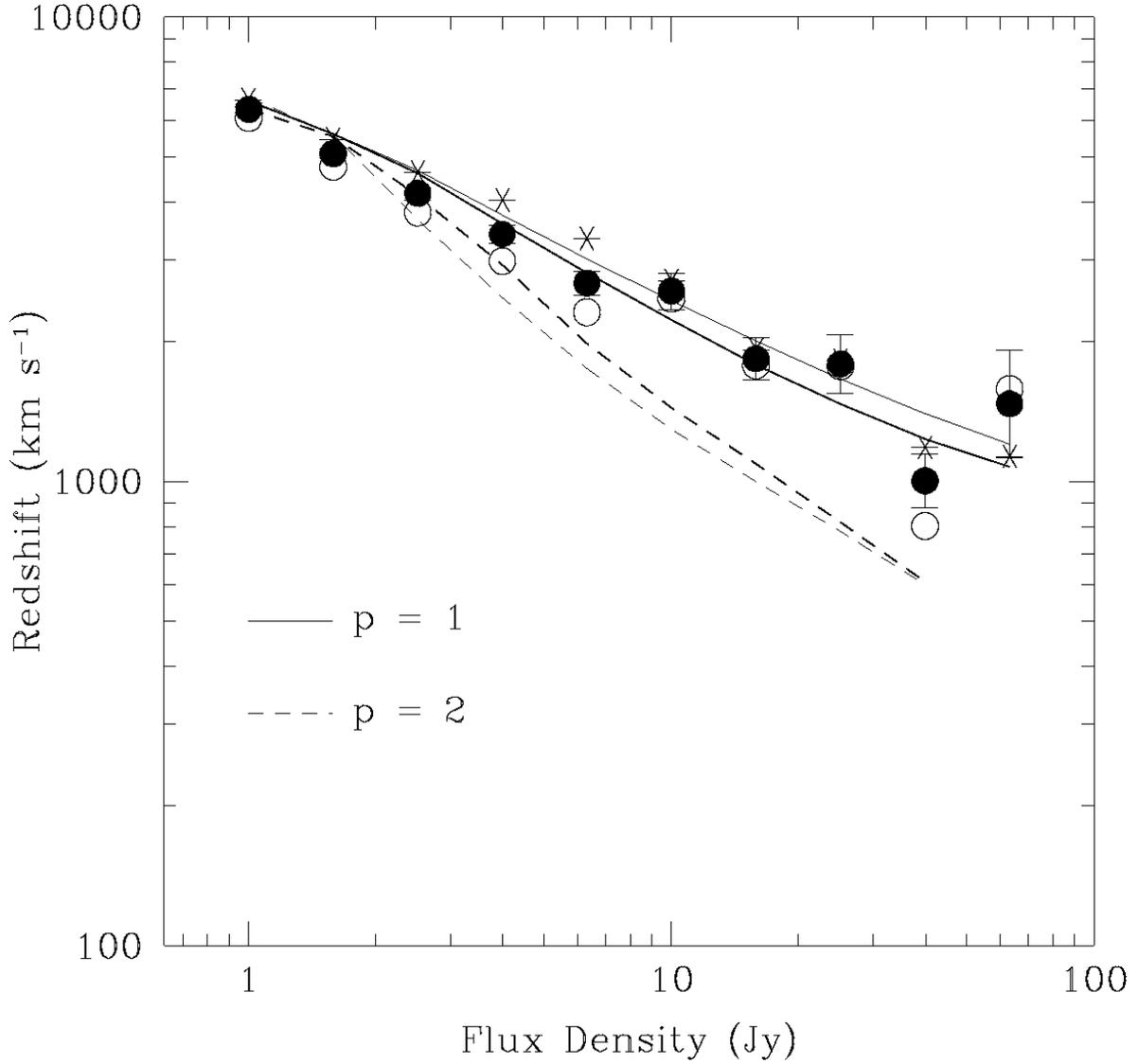}
\caption[]{The solid points give the mean log redshift of galaxies in bins
of log flux density from the entire \iras\ 1.2 Jy sample, averaging
over galaxies with redshifts between 500 and 20,000 \kms.  The error bars are
the error in the mean (that is, the standard deviation divided by the
square root of the number of points in each bin).  The open circles are
the mean log redshift for galaxies in the Northern Galactic Hemisphere,
while the stars are for the Southern Galactic Hemisphere.  The light solid
line gives the expected curve assuming $p = 1$ and a homogeneous
universe, while the light dashed line assumes $p = 2$.  The heavy
solid and dashed lines give the expected density field taking into
account the density field found non-parametrically in
Fig.~\ref{fig:zdist}. 
\protect\label{fig:Soneira}}
\end{figure}

  A final approach is suggested by Soneira (1979) and Cho\l oniewski
(1995); cf., Nicoll \& Segal (1982).   The expectation value of $\log z$ as
a function of flux density follows directly from Eq.~(\ref{eq:z-given-L}):
\begin{equation} 
\vev{\log z|f} = {\int \d z\,  {\d V \over \d z}\, \log z\, \f(L,z) \over \int \d z\, {\d V
\over \d z}\, \f(L,z)} = 
 {\int \d z\,  \log z\, z^{3/p -1}\rho(z) \Phi(L) \over \int \d z\, z^{3/p - 1}
\rho(z) \Phi(L)}\qquad,
\label{eq:z-given-f} 
\end{equation}
where $L = 4\,\pi (cz/h_p)^{2/p} \nu f$.  Unlike the expressions for
the conditional probabilities, neither $\rho$ nor $\Phi$ drops out of
the expression. 
Fig.~\ref{fig:Soneira} shows this statistic in bins of log flux density
for the full \iras\ sample (solid points, with errors in the mean
shown) and also for the Northern and Southern Galactic hemispheres of
the sample separately (open circles and stars, respectively), as a
test of the robustness of the statistic to density inhomogeneities.
The averaging is done for galaxies in the redshift range 500 \kms\ to
20,000 \kms.  The $p = 1$ prediction assuming $\rho(z) = 1$ is given
as the light solid line, while $p = 2$ is shown with the light dashed
line.  These curves are not pure power laws, because of the upper
limit on redshift we imposed.  The $p = 1$ line is not a perfect fit
to the data; one is presumably seeing the residual effect of density
inhomogeneities, as is made clear by the differences between the
Northern and Southern hemispheres, especially at large flux densities.
However, $p = 2$ does much more poorly, especially at large flux
densities.  This is not unexpected: as Fig.~\ref{fig:zdist} showed, the homogeneity
assumption is a very poor one for the $p = 2$ model.  We can include
this effect by
carrying out the integration of Eq.~(\ref{eq:z-given-f}), using the
density field found non-parametrically in Fig.~\ref{fig:zdist}; the
results are shown as the heavy solid and dashed line for $p = 1$ and
$p = 2$, respectively.  The $p = 1$ line is now in excellent agreement
with the data, while the $p = 2$ line approaches the data points
somewhat more closely, but is still far from a good fit to the data.
It is not clear whether this represents a breakdown of the universal
luminosity function assumption which went into
Eq.~(\ref{eq:z-given-f}), for the case of $p = 2$. 


\placefigure{fig:Soneira}

\section{Conclusions}

The density-dependent methods of comparing the cosmologies favor $p=1$
quite unambiguously.  However, one reaches this conclusion only with
the auxiliary assumption that the density field of galaxies approaches
homogeneity on large scales. In the case of the density-independent
comparisons of the distribution of flux density, we have recourse to
quantitative $\chi^2$ testing of goodness-of-fit, and we find that the
standard $p=1$ cosmology yields better agreement than does the $p=2$
case.  But perhaps most surprising is the extent to which these
methods do a poor job of discriminating between different values of
$p$, implying that the results of any analysis predicated on their use
should be interpreted with great caution.  These results are in accord
with the analytic arguments of Petrosian (1993): without additional
assumptions, questions of cosmology and the appropriate luminosity
function cannot be decoupled from flux density and redshift data, and
that one cannot determine both simultaneously.

A further methodological pitfall which affects SNWZ is the sensitivity
of the non-parametric luminosity function to bin size. SNWZ used
very coarse binning and therefore were working with luminosity
functions inaccurate at the faint end.  Since this effect is less
pronounced for $p=2$ than for $p=1$, the luminosity function that SNWZ
employed for $p=2$ was less in error than the one they employed for
$p=1$.  We found that with equally coarse binning, this effect causes
$p = 2$ to be favored, both by the statistics used by SNWZ and by our
own comparisons of the flux density distribution.  

Therefore from redshift survey data alone, one can conclude that $p =
1$ is preferred only weakly over $p = 2$ if we only allow ourselves the assumption of a
universal luminosity function.  When one makes the additional
assumption that the distribution of galaxies approaches isotropy on
large scales, the case for $p = 1$ becomes {\it much\/} stronger;
indeed, if $p = 2$, one would need to argue that the density field
of galaxies beyond 10,000 \kms\ drops dramatically and steadily with
redshift.  When this is combined with the observed linearity of the
redshift-distance diagram using measurements of extragalactic standard
candles (cf., Mould \etal\ 1991; Lauer \& Postman 1992; Hamuy \etal\
1995, 1996; Perlmutter \etal\ 1996; Riess \etal\ 1996), the evidence for the
Hubble law becomes overwhelming.

\acknowledgments

We thank Dr.~I. Segal for detailed comments on an earlier draft of
this paper, and Marc Davis for suggesting we include the Soneira
(1979) statistic in this paper.  D.M.K. was supported by a National
Science Foundation Graduate Fellowship.  M.A.S. acknowledges support
from the W.M. Keck Foundation, NASA Theory Grant NAG5-2882, and the
Alfred P. Sloan Foundation. 

\appendix
\section{Interpolation of the piecewise-constant luminosity function}

\begin{figure}
\plotone{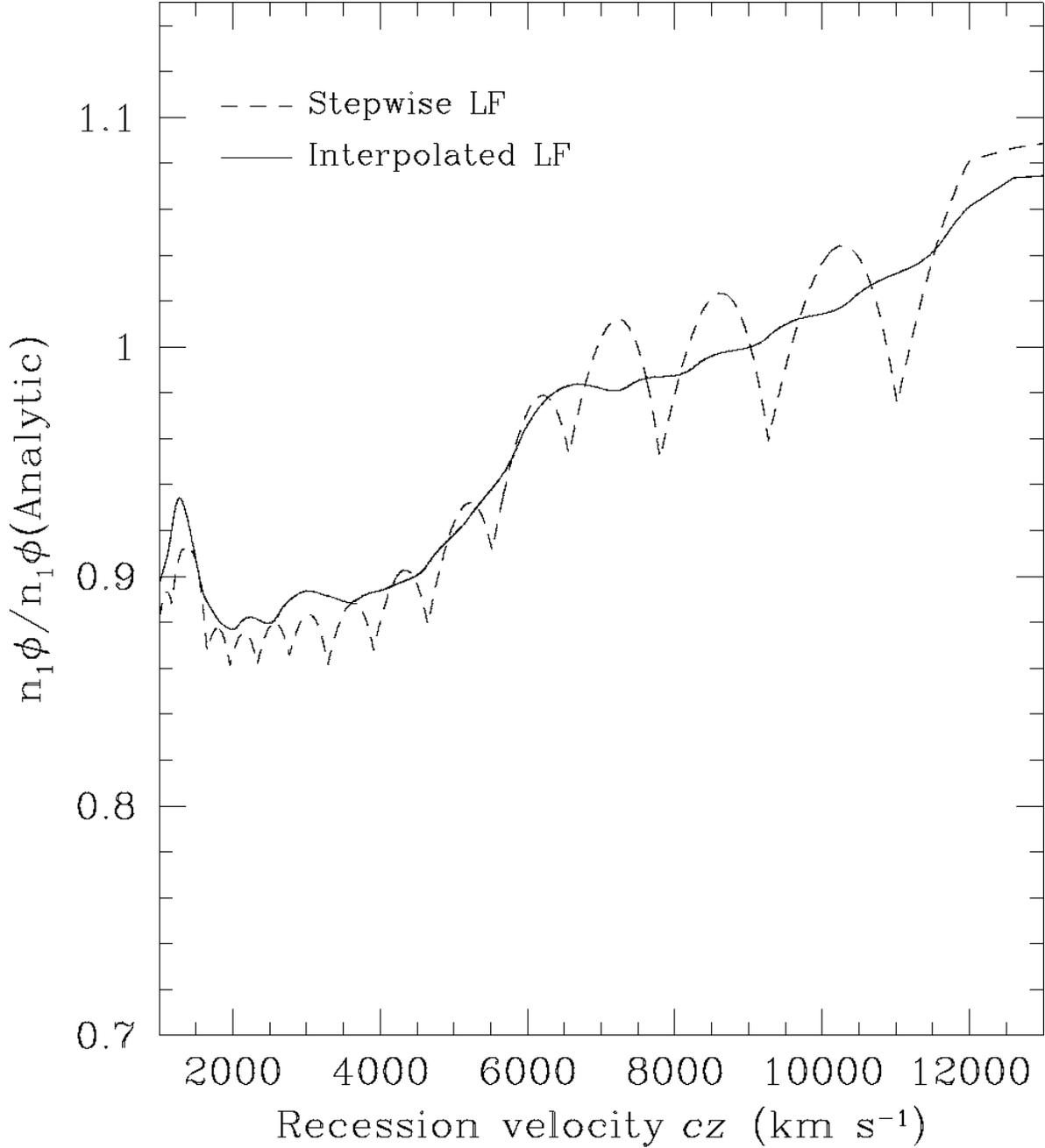}
\caption[]{The solid line shows the ratio of the normalized selection
function $n_1\phi(z)$ calculated from the interpolated luminosity
function described in the Appendix, to that for the
analytic luminosity function of Eqs.~(\ref{eq:Yahil}) and
(\ref{eq:Yahil-diff}). The dashed line is the ratio of $n_1 \phi(z)$
for the stepwise luminosity function to that of the analytic
luminosity function. Notice the scalloping in the
latter. \protect\label{fig:scallop}}
\end{figure}

The piecewise-constant luminosity function as defined by Efstathiou
\etal\ (1988) has one drawback; the corresponding selection
function (defined from Eqs.~\ref{eq:selfunct} and \ref{eq:Psi}) shows a
scalloping effect (cf., Strauss \& Koranyi 1994). This is illustrated
in Fig.~\ref{fig:scallop}, which shows the ratio of the selection function
calculated from the piecewise-constant luminosity function (dashed line)
calculated with bins of $\Delta \log L = 0.15$, to the analytic
selection function (from Eq.~\ref{eq:Yahil}).  These were calculated
for $p = 1$.  The selection function
of the piecewise-continuous luminosity function has a discontinuity in
slope at the edge of each bin (as indeed it must; the slope is
proportional to the luminosity function itself). The effect is
minimized here by using small bins, but nevertheless it would be best
to eliminate this altogether. We do so by generalizing the method of Efstathiou
\etal\ (1988) to define the luminosity function not as constant in a
series of bins, but rather as a series of line segments connecting
bins. The formalism of Efstathiou \etal\ can be carried over exactly,
if we replace Equations 2.8-2.11 of that paper with the following.
Define the luminosity function to be a series of line segments
connecting a series of points $(L_k, \Phi_k), k = 1, \ldots,
n$\footnote{Unlike Nicoll \& Segal (1983), we do not explicitly split our
calculation of $\Phi$ into those regions above and below
$L_{min}(z_{max})$.  There is therefore no reference to $n_{ll}$ in
this Appendix.}; thus
\begin{equation}
\Phi(L) = \sum_{k = 1}^{n} \Phi_k W(L,k)\qquad .
\label{eq:phi-def} 
\end{equation}
Note
that in the present paper, we refer to the luminosity function as
$\Phi$, while Efstathiou \etal\ refer to it as $\phi$.  The logarithm
of the likelihood is then given by 
\begin{equation} 
\ln {\cal L} = \ln\left[\sum_{i = 1}^N \sum_{k = 1}^{n}\Phi_k
W(L,k)\right] - \ln \left[\sum_{i = 1}^N \sum_{k =
1}^{n}\Phi_k[H(L_{min}(z_i),k) - H(L_{max}(z_i),k)] \right]\qquad ,
\label{eq:loglike-step} 
\end{equation}
where $N$ is the total number of galaxies in the sample. 
Note that Eq.~(2.9) of Efstathiou \etal\ is erroneously missing the
summation over $k$ in the first term on the right-hand side. Here,
$L_{min}$ is defined in Eq.~(\ref{eq:lmin}), and $L_{max}(z_i) = 4 \pi
(cz/h_p)^{2/p} \nu f_{max}$ if one derives the luminosity function for
samples with an upper limit on flux density.  If there is no upper limit on
flux density, $ H(L_{max}(z_i),k) = 0$.

The expressions for $W$ and $H$ are more complicated than those in 
Efstathiou \etal\ (1988), but they just involve linear interpolation: 
\begin{equation} 
{W(L,k) = \cases{
{{L - L_{k - 1}}\over{L_k - L_{k-1}}} & $L_{k - 1} < L \le L_k$ and
$k \ne 1$\cr
{{L_{k+1} - L}\over{L_{k+1} - L_k}} & $L_k < L \le L_{k+1}$ and $k
\ne n$\cr
0&otherwise.\cr
}}
\label{eq:W} 
\end{equation}
\begin{equation} 
{H(L,k) = \cases{
{1 \over 2} (L_{k+1} - L_{k - 1}) & $L \le L_{k - 1}$ and $k \ne 1$\cr
{1 \over 2} \left[L_{k+1} - L_k + (L_k + L - 2L_{k - 1}){{L_k
-L}\over{L_k - L_{k - 1}}}\right] & $L_{k - 1} < L \le L_k$ and $k \ne 1$\cr
{1 \over 2} {{(L_{k+1} - L)^2}\over{L_{k+1} - L_k}} & $L_k \le L <
L_{k+1}$ and $k \ne n$\cr
0&otherwise.\cr}}
\label{eq:H} 
\end{equation}
Note that here, we define $L_{n + 1} \equiv L_{n}$. 
Finally, the integral constraint we use is:
\begin{equation} 
\sum_k \Phi_k (L_k - L_{k - 1}) = 1\qquad.
\label{eq:constraint}
\end{equation}
One can normalize after this using the quantity $n_1$ defined in
Eq.~(\ref{eq:n1}) (as indeed we have done in Fig.~\ref{fig:scallop}).
The solid line in Fig.~\ref{fig:scallop} shows the ratio of selection
function for the resulting continuous luminosity function to that for
the analytic luminosity function; the scalloping effect has gone
away. Other than the scalloping effect, the interpolated selection
function, and that using the piece-wise continuous method are in good
agreement, implying that this interpolation technique has very little
effect on the derived luminosity function (although both differ at the
10\% level from the analytic luminosity function; this is a residual
effect of the finite binning, as described in \S 2).  However, it does
have a non-negligible effect on the derived selection function, and we
have used it in all calculations requiring a nonparametric luminosity
function in this paper.
\placefigure{fig:scallop}
\clearpage
 
\begin{deluxetable}{ccccc}
\footnotesize
\tablecaption{Parameters of the parameterized luminosity function. \label{table:yahil123parms}}
\tablewidth{0pt}
\tablehead{
\colhead{$p$} & \colhead{$\alpha$} & \colhead{$\beta$} &
\colhead{$\log _{10} (L_*)$} &\colhead
{$n_1$}\\&&&\colhead{$L_\odot$}&\colhead{($10^{-6}(\kms)^{-3/p}$)}}
\startdata
1 & \phs$0.49\pm0.07$ & 1.81$\pm$0.12 & $9.68^{+0.10}_{-0.13}$ & 0.058 \nl
2 & $-0.02\pm0.14$ & 2.47$\pm$0.14 & $7.98^{+0.10}_{-0.13}$ & 4.357 \nl
3 & $-0.81\pm0.30$ & 3.07$\pm$0.24 & $7.16^{+0.10}_{-0.14}$ & 25.25 \nl
 
\enddata

\end{deluxetable}

\begin{deluxetable}{lrrrrrrl}
\footnotesize
\tablecaption{Statistics of SNWZ for the sample with $2000 < cz <
20,000 \kms$.\label{table:segalstats}}
\tablewidth{0pt}
\tablehead{Model&$n_{ll}$\tablenotemark{a}&$n$\tablenotemark{b}&$\Delta \log
L$\tablenotemark{c}&$s$\tablenotemark{d}&
$\beta$\tablenotemark{e}&$\chi^{2\,}$\tablenotemark{f}&{}}
\startdata
Real data & --- & --- & --- &0.628 & 0.610 & --- \nl
$p=1$ &  9  & 13 & 0.365 & 0.686 & 0.700 & 377.7 &Bin size used by SNWZ \nl
$p=1$ & 10  & 14 & 0.302 & 0.661 & 0.677 & 203.0 & \nl
$p=1$ & 25  & 37 & 0.109 & 0.599 & 0.591 & 10.7 & \nl
$p=1$ & --- & ---&  ---  & 0.587 & 0.536 &  5.1   &Parameterized Luminosity Function \nl
$p=2$ &  8  & 18 & 0.187 & 0.616 & 0.520 & 115.7  &Bin size used by SNWZ \nl
$p=2$ & 10  & 21 & 0.137 & 0.604 & 0.489 & 73.0 & \nl
$p=2$ & 25  & 53 & 0.052 & 0.592 & 0.452 & 40.4 & \nl
$p=2$ & --- & ---&  ---  & 0.588 & 0.444 & 38.9   &Parameterized Luminosity Function \nl
$p=3$ &  7  & 22 & 0.131 & 0.613 & 0.369 & 103.9   &Bin size used by SNWZ \nl
$p=3$ & 10  & 29 & 0.089 & 0.608 & 0.349 & 76.7 & \nl
$p=3$ & 25  & 71 & 0.034 & 0.604 & 0.332 & 59.4 \nl
$p=3$ & --- & ---&  ---  & 0.597 & 0.328 & 71.8   &Parameterized Luminosity Function \nl
\enddata
\tablenotetext{a}{The number of bins in the luminosity function below
$L_{min}(z_{max})$.}
\tablenotetext{b}{The total number of bins in the luminosity function.}
\tablenotetext{c}{The size of the bins in $\log L$.}
\tablenotetext{d}{The standard deviation of the apparent magnitudes.}
\tablenotetext{e}{The slope of the apparent magnitude regression on $\log
cz$.}
\tablenotetext{f}{The $\chi^2$ of the predicted flux density distribution of the
sample to that observed.}
\end{deluxetable}



\begin{deluxetable}{lrrrrrrrr}
\footnotesize
\tablecaption{$\chi^2$ analysis for flux density distribution of
galaxies, $n_{ll} = 25$. 
   \label{table:fluxchi2}}
\tablewidth{0pt}
\tablehead{&&&&$p = 1$&&$p = 2$&&$p = 3$\\$cz$&$\nu$&\# of Galaxies&$\chi^2$&$\chi^2/\nu$&
$\chi^2$&$\chi^2/\nu$&$\chi^2$&$\chi^2/\nu$\\
\kms}
\startdata
 500--2000 &    19 &  674 &  23.1 &       1.21 &       22.0 &       1.16&       39.8 &       2.09\nl
2000--4000 &    17 &  901 &  14.2 &       0.84 &       20.5 &       1.21&       23.8 &       1.40\nl
4000--6000 &    15 & 1068 &  13.8 &       0.92 &       17.0 &       1.14&       19.9 &       1.32\nl
6000--8000 &    13 &  697 &  18.5 &       1.42 &       26.4 &       2.03&       32.1 &       2.47\nl
8000--12,000&   11&  878 &   5.0 &       0.46 &        8.6 &       0.78&       17.6 &       1.60\nl
12,000--20,000& 10 &  639 &   9.8 &       0.98 &       16.6 &       1.66&       24.9 &       2.49\nl
Entire Sample&  19 & 4857 &  12.2 &       0.64 &       45.7 &       2.40&       70.4 &       3.70\nl
\enddata     
\end{deluxetable}

\begin{deluxetable}{lrrrrrrrr}
\footnotesize
\tablecaption{$\chi^2$ analysis for flux density distribution of
galaxies, using SNWZ binning.
   \label{table:fluxchi2_segalstep}}
\tablewidth{0pt}
\tablehead{&&&&$p = 1$&&$p = 2$&&$p = 3$\\$cz$&$\nu$&\# of Galaxies&$\chi^2$&$\chi^2/\nu$&
$\chi^2$&$\chi^2/\nu$&$\chi^2$&$\chi^2/\nu$\\
\kms}
\startdata
 500--2000 &   19 &  674 &         41.8 &       2.20 &       23.9 &       1.26&       35.1 &       1.85\nl
2000--4000 &   17 &  901 &         84.8 &       4.99 &       31.2 &       1.84&       25.1 &       1.48\nl
4000--6000 &   15 & 1068 &        119.7 &       7.98 &       33.9 &       2.26&       25.7 &       1.71\nl
6000--8000 &   13 &  697 &        109.2 &       8.40 &       57.2 &       4.40&       45.9 &       3.53\nl
8000--12,000&   11 &  878 &         61.6 &       5.60 &       22.1 &       2.01&       26.4 &       2.40\nl
12,000--20,000& 10 &  639 &        19.5 &       1.95 &       19.6 &       1.96&       28.2 &       2.82\nl
Entire Sample& 19 & 4857 &        299.6 &      15.77 &      111.3 &       5.86&      106.9 &       5.63\nl
\enddata
\end{deluxetable}

\clearpage

\end{document}